\documentclass{ws-ijmpa}

\begin{document}

\markboth{Geng and Hsiao}
{Direct $CP$ and  $T$ Violation in Baryonic B Decays}

%
\catchline{}{}{}{}{}
%
\def\ra{\rightarrow}
\def\be{\begin{eqnarray}}
\def\ee{\end{eqnarray}}
\def\lln{\left<}
\def\rln{\right>}
\def\gam{\gamma}
\def\prl{Phys. Rev. Lett.~}
\def\pr{Phys. Rev.~}
\def\pl{Phys. Lett.~}
\def\np{Nucl. Phys.~}
\def\prl{Phys. Rev. Lett.~}
\def\tp{\vec v_1 \cdot (\vec v_2 \times \vec v_3)}

\title{Direct $CP$ and  $T$ Violation in Baryonic B Decays
  }

\author{  C.Q. GENG$^{1}$
  and Y.K. HSIAO$^{2}$}

\address{$^{1}$Department of Physics, National Tsing-Hua University,
Hsinchu, Taiwan 300 \\
$^2$Institute of Physics, Academia Sinica, 
Taipei, Taiwan 115\\
}

\maketitle


\begin{abstract}
We review the direct CP and T violation in the three-body baryonic $B$ decays
 in the standard model.
 In particular, we emphasize that
the direct CP violating asymmetry in
$B^\pm\to p\bar p K^{*\pm}$ is around 22$\%$
and  
the direct $T$ violating asymmetry in
$\bar B^0 \ra \Lambda \bar p \pi^+$ 
can be as large as
$12\%$, which are accessible to the current B factories at KEK
and SLAC as well as SuperB and LHCb.

\end{abstract}

\vspace{1.cm}
Direct CP violation has been measured in both $K^0$ and $B^0$ systems \cite{pdg},
but it has not been observed and conclusive in $K^\pm$ and $B^\pm$ systems \cite{pdg}, respectively.
On the other hand,
  T violation has been only seen in the $K^0$ process \cite{pdg}, 
  related to the indirect CP violating parameter $\epsilon_K$, 
  whereas no T violating effect has been found in either  $K^\pm$ or $B$ systems yet.
 In the standard model (SM), it is clear that
   the unique phase   
 of the
Cabbibo-Kobayashi-Maskawa (CKM) matrix \cite{CKM}  is responsible for both observed CP and T violating effects.
In this talk, we would like to explore the possibility to detect
the direct CP and T violation in the $B$ systems in the current B-factories as well as the future ones such as
SuperB and LHCb. In particular,
we concentrate on the three-body charmless baryonic
processes. Our goal of the talk is to test the CKM
paradigm of CP violation and unfold new physics.

In the framework
  of local quantum field theories, 
  T-violation implies CP-violation (and vice versa), because of the
  CPT invariance of such theories. 
Moreover, 
  no violation of CPT symmetry has been found~\cite{pdg}.
  Still, it will be worthwhile to
  remember that outside this framework of local quantum field
  theories, there is no
  reason for the two symmetries to be linked~\cite{CPT}. 
  Therefore, it would be
  interesting to directly investigate T violation in B decays,  rather than inferring
  it as a consequence of CP-violation.
The characteristic observables of the direct CP and T violation are rate asymmetries
and momentum correlations, respectively. For example,
in (conjugate) processes such as $B\to {\bf B}\bar{\bf B'}M$ ($\bar{B}\to {\bf\bar{ B}}{\bf B'}\bar{M}$),  
the direct
CP asymmetry
 arises if both the weak ($\gamma$) and strong
($\delta$) phases are non-vanishing, given by
\be 
A_{CP}&=&{\Gamma (B\to {\bf B}\bar{\bf B'}M)-\Gamma (\bar{B}\to {\bf\bar{ B}}{\bf B'}\bar{M})
\over \Gamma (B\to {\bf B}\bar{\bf B'}M)+\Gamma (\bar{B}\to {\bf\bar{ B}}{\bf B'}\bar{M})}
 \propto \sin
\gamma \sin \delta\,,
 \label{cpa} 
\ee
 whereas the direct T violation is related to 
 the correlations known as triple product
correlations (TPC's), such as $\vec{s}_{\bf B}\cdot (\vec{p}_{\bf B}\times \vec{p}_M)$, given 
by
\be
 {\cal A}_T &=& {1 \over 2}(A_T-\bar A_T)
\propto \sin \gamma \cos \delta.
 \ee
 where \cite{KT}
 \be A_T = {\Gamma (\vec{s}_{\bf B}\cdot (\vec{p}_{\bf B}\times \vec{p}_M) > 0)
- \Gamma (\vec{s}_{\bf B}\cdot (\vec{p}_{\bf B}\times \vec{p}_M) < 0) \over 
\Gamma (\vec{s}_{\bf B}\cdot (\vec{p}_{\bf B}\times \vec{p}_M) > 0)
+\Gamma (\vec{s}_{\bf B}\cdot (\vec{p}_{\bf B}\times \vec{p}_M) < 0)
}\;,
\label{atp}
\ee
and $\bar{A}_T$ is the corresponding asymmetry of the conjugate process.
It is interesting to note that to have a non-zero value of $A_{CP}$, both weak and strong phases are needed,
whereas in the vanishing limit of the strong phase,  ${\cal A}_T$ is maximal.
 Furthermore, there is no contribution \cite{KT} to
${\cal A}_T$  from final state interaction due
to electromagnetic interaction.

 From the effective Hamiltonian at the quark level for $B$ decays
\cite{Hamiltonian}, the amplitudes of $B^-\to p\bar p K^-$ and
$B^-\to p\bar p K^{*-}$ are approximately given by
\cite{ChuaHouTsai2,GH05,AngdisppK,ppKstar}
\begin{eqnarray}\label{eq1}
{\cal A}_K&\simeq&i\frac{G_F}{\sqrt 2}m_b f_K\bigg[\alpha_K\langle
p\bar p|\bar u b|B^-\rangle+\beta_K\langle p\bar p|\bar u\gamma_5
b|B^-\rangle\bigg]\,, \nonumber\\
{\cal A}_{K^*}&\simeq&\frac{G_F}{\sqrt
2}m_{K^*}f_{K^*}\varepsilon^{\mu}\alpha_{K^*}\langle p\bar p|\bar
u\gamma_\mu(1-\gamma_5) b|B^-\rangle\;,
\end{eqnarray}
respectively,
where $G_F$ is the Fermi constant,
 $f_{K^{(*)}}$ is the meson decay constant, given by
$\langle K^-|\bar s\gamma_\mu \gamma_5 u|0\rangle=-if_K q_\mu$
($\langle K^{*-}|\bar s\gamma_\mu u|0\rangle
=m_{K^*} f_{K^*}\varepsilon_\mu$)
with $q_\mu$ ($\varepsilon_\mu$) being the four momentum (polarization) of $K^-$ ($K^{*-}$), and $\alpha_{K^{(*)}}$ and $\beta_{K}$ are defined by
\begin{eqnarray}\label{eq2}
\alpha_K(\beta_K)&\equiv& V_{ub}V_{us}^*a_1-V_{tb}V_{ts}^*\bigg[a_4\pm a_6\frac{2 m_K^2}{m_b m_s}\bigg]\;,
\nonumber\\
\alpha_{K^*}&\equiv& V_{ub}V_{us}^*a_1-V_{tb}V_{ts}^*a_4\;,
\end{eqnarray}
where $V_{ij}$ are the CKM matrix elements
and $a_i$ ($i=1,4,6$) are given by
\begin{eqnarray}\label{a146}
a_1=c_1^{eff}+\frac{1}{N_c}c_2^{eff}\;,\;a_4=c_4^{eff}+\frac{1}{N_c}c_3^{eff}\;,\;a_6=c_6^{eff}+\frac{1}{N_c}c_5^{eff}\;,
\end{eqnarray}
with $c_i^{eff}\;(i=1,2, ..., 6)$ being effective Wilson
coefficients (WC's) shown in Ref.~\cite{Hamiltonian} and $N_c$
the  color number for the color-octet terms.  We note that
for the decay amplitudes in Eq. (\ref{eq1}) we have neglected
the small contributions \cite{AngdisppK,ppKstar} from
$\langle
p\bar p|J_1|0\rangle\langle K^{(*)}|J_2|B\rangle $  
involving the 
 $vacuum \to p\bar p$  time-like  
baryonic form factors~\cite{NF}, where
$J_{1,2}$ can be (axial-)vector or (pseudo)scalar currents.
However, 
in our numerical analysis
we will keep all amplitudes including the ones neglected in 
Eq. (\ref{eq1}).
Numerically, the CKM parameters are taken to be
\cite{pdg} $V^{\;}_{ub}V^{*}_{us}=A\lambda^4(\rho-i\eta)$ and
$V^{\;}_{tb}V^{*}_{ts}=-A\lambda^2$ with $A=0.818$,
$\lambda=0.2272$, the values of $(\rho,\eta)$ are $(0.221,0.340)$~\cite{pdg}. 
We remark that $a_i$  contain both weak and strong
phases, induced by $\eta$ and quark-loop rescatterings. Explicitly, at the scale $m_b$
and $N_c$=3, we obtain a set of $a_1$, $a_4$, and $a_6$ as
follows:
\begin{eqnarray}\label{set1}
a_1&=&1.05\;,\nonumber\\
a_4&=&\big[(-427.8\mp 9.1\eta-3.9\rho)+i(-83.2\pm 3.9\eta-9.1\rho)\big]\times
10^{-4} \;,\nonumber\\
a_6&=&\big[(-595.5\mp 9.1\eta-3.9\rho)+i(-83.2\pm 3.9\eta-9.1\rho)\big]\times
10^{-4}\;,
\end{eqnarray}
for the $b\to s$ ($\bar b\to \bar s$) transition.

 From Eq. (\ref{eq1}),
we derive the simple results for the direct CP asymmetries of the $K^{(*)}$ modes as 
follows:
\begin{eqnarray}\label{Acp2}
A_{CP}(K^{(*)})&\simeq &\frac{|\alpha_{K^{(*)}}|^2-|\bar
\alpha_{K^{(*)}}|^2}{|\alpha_{K^{(*)}}|^2+|\bar \alpha_{K^{(*)}}|^2}\,,
\end{eqnarray}
where
 $\bar\alpha_{K^{(*)}}$  denote the
values of the corresponding antiparticles.
It is easy to see that
$A_{CP}(K^{(*)})$ are independent of the phase spaces as well as the
hadronic matrix elements. As a result, the hadron parts along with
their uncertainties in $A_{CP}(K^{(*)})$ are divided out in Eq.
(\ref{Acp2}). We note that the CP asymmetries in Eq. (\ref{Acp2}) are related to
the weak phase of $\gamma (\phi_{3})$ \cite{pdg}.

Our results on the direct CP violation are summarized
in Table \ref{pre}.
\begin{table}[htbp]
\tbl{Direct CP asymmetries in $B\to p\bar{p} M$.}
{\begin{tabular}{|c|c|c|c|c|}
\hline 
$A_{CP}( M)$
&$A_{CP}(K^{*\pm})$
&$A_{CP}(K^\pm)$
&$A_{CP}( K^{*0})$
&$A_{CP}(\pi^\pm)$
\\\hline
Our work \cite{ppKstar}&0.22&0.06&0.01&-0.06\\\hline
BaBar \cite{CPppK}&$0.32\pm0.14$&$-0.13^{+0.09}_{-0.08}$&$0.11\pm0.14$ &$0.04\pm0.08$
\\\hline
Belle \cite{Wang1}& &$-0.02\pm0.05$&&$-0.17\pm0.10$
\\\hline
\end{tabular}
\label{pre}}
\end{table}
In the table, we have included the current experimental data as well as the decay modes of
$B^\pm\to p\bar{p} \pi^\pm$.
We note that 
the possible
fluctuations induced from non-factorizable effects, time-like
baryonic form factors and CKM matrix elements for 
$A_{CP}(K^{(*)})$ are about $0.01$ (0.04), 0.003 (0.01) and 0.01 (0.01), respectively. 
The uncertainties from time-like baryonic form factors  are constrained by the data of $\bar B^0\to
n\bar p D^{*+}$ and $\bar B^0\to\Lambda \bar p \pi^+$ \cite{NF}
and the errors on the CKM elements are from $\rho$ and $\eta$
 given in Ref. \cite{pdg}.
It is interesting to point out that
the large value of $A_{CP}(B^\pm\to p\bar p
K^{*\pm})$=22\% is in agreement with  the BABAR data of $(32\pm 14)\%$. 
However, taken at face value; the sign of our prediction  
$A_{CP}(B^{\pm}\to p\bar p K^\pm)$ is different from those
by BABAR \cite{CPppK} and BELLE \cite{Wang1}
  Collaborations.
 Since the uncertainties of both experiments are still large
it is too early to make a firm conclusion.

 For the direct  T violation in the three-body charmless baryonic B decays \cite{GHT1}.
  we  concentrate on  $\bar B^0 \ra \Lambda \bar p \pi^+$
by  looking for the TPC of
the type $\vec s_{\Lambda} \cdot (\vec p_{\bar p}\times\vec
p_{\Lambda})$. 
It is interesting to note that~\cite{belle0}: 
\be\label{exbr}
 Br( B^0 \ra \bar{\Lambda} p\pi^-) =
(3.29\pm0.47) \times 10^{-6}\,\gg\,
Br(B^- \ra\Lambda \bar p) &<& 4.6 \times 10^{-7}\,.
\ee 
The enhancement of three-body decay over the two-body one is
due to the reduced energy release in $B$ to $\pi$ transition by the
fastly recoiling $\pi$ meson that favors the dibaryon production
\cite{soni}. Theoretical estimations baryonic B decays are made
\cite{GH05,chua,group}, in consistent with
the experimental observations.

In the factorization method, the decay amplitude of $\bar B^0 \to
\Lambda \bar p \pi^+$ contains the $\bar B^0\to\pi^+$ transition
and $\Lambda \bar p$ baryon-pair inducing from the vacuum. The
contributions to the decay at the quark level are mainly from
$O_1$, $O_4$ and $O_6$ operators.
  From these operators and the factorization
approximation, the decay amplitude is given by \cite{GHT1,chua}
\begin{eqnarray}\label{M146}
M&=&M_1+M_4+M_6\nonumber\;,\\
M_i&=& {G_f \over \sqrt{2}} \lambda_i a_i \lln \pi^+| \bar u
\gamma^\mu(1-\gamma_5)b|\bar B^0\rln \lln \Lambda \bar p|\bar s
\gamma_\mu(1-\gamma_5)u|0\rln
\label{m1}\;,\ (i=1,4)\,,\nonumber\\
M_6 &=& {G_f \over \sqrt{2}} V_{tb} V_{ts}^* 2a_6 \lln \pi^+ |
\bar u \gamma^\mu(1-\gamma_5)b|\bar B^0 \rln {(p_\Lambda +p_{\bar
p})_\mu \over {m_b - m_u}} \lln \Lambda \bar p|\bar s
(1+\gamma_5)u|0\rln \label{m6}\;,
\end{eqnarray}
where $\lambda_1=V_{ub} V_{us}^*$, $\lambda_4=-V_{tb} V_{ts}^*$
and $a_i$  are defined in Eq. (\ref{a146}).
 From Eq. (\ref{m6}),
the T-odd transverse polarization asymmetry
$P_T$ is found to be
\begin{eqnarray}
\label{AT}
P_T& \propto &
\bigg(V\cdot S-A\cdot P\bigg)Im\left(V^{\;}_{ub}V^{*}_{us}V^{\;}_{tb}V^{*}_{ts}a_1^{\;}a_6^*\right)
\,,
\end{eqnarray}
where $S,P,V$ and $A$ are combinations of form factors, 
given by~\cite{GHT1}
\begin{eqnarray}\label{ABFG}
V&=&F^{B \to \pi}_1(t)[F_1(t)+F_2(t)]\;,\ 
A\;=\;F^{B \to \pi}_1(t)g_A(t)\;,\nonumber\\
S&=&\frac{m_B^2-m_\pi^2}{m_b-m_u}F^{B \to \pi}_0(t)f_S(t)\;,\
P\;=\;\frac{m_B^2-m_\pi^2}{m_b-m_u}F^{B \to \pi}_0(t)g_P(t)\;.
\end{eqnarray}
It is noted that the $V\cdot S$ ($A\cdot P$) term is from
vector-scalar (axialvector-pseudoscalar) interference and there is
no T-odd term from $Re(M_1 M_4^\dagger)$ due to the same current
structures.
 In Eq. (\ref{ABFG}),
$F_{1,0}^{B\to\pi}(t)$ are the well known mesonic $\bar{B}^0\to \pi^+$ transition form factors  \cite{MS},
while $F_{1,2}(t)$, $g_A(t)$, $h_A(t)$, $f_S(t)$ and $g_P(t)$ are the  $0\to \Lambda\bar{p}$
time-like baryonic  form factors, defined in Ref. \cite{GHT1}.
Based on the QCD counting rules \cite{Brodsky1} and $SU(3)$ flavor symmetry, 
at $t\to \infty$ one has that
\be
F_1(t)+ F_2(t)\sim g_A(t)\sim h_A(t)\sim f_S(t)\sim g_P(t)\sim {C\over t^2}\,.
\ee
In this limit $A_T\to 0$ and thus no T violation is expected. 
However, at the finite $t$ there are some high power terms of  the $t$ expansion.
A simple scenario of the power expansions for the baryonic form factors is as follows:\cite{GH-f}
\be
\label{SS}
F_1(t)+ F_2(t)=\left({C\over t^2}+{D\over t^3}\right)\left[\ln\left({t\over \Lambda_0^2}\right)\right]^{-\gamma}
\,, &&
g_A(t)=\left({C\over t^2}\right)\left[\ln\left({t\over \Lambda_0^2}\right)\right]^{-\gamma}\,,\
\nonumber\\
f_S(t)=n_q\left({C\over t^2}+{D\over t^3}\right)\left[\ln\left({t\over \Lambda_0^2}\right)\right]^{-\gamma}\,,
&&
g_P(t)=n_q\left({C\over t^2}\right)\left[\ln\left({t\over \Lambda_0^2}\right)\right]^{-\gamma},\ \ \
\ee
where $n_q=(m_\Lambda-m_p)/(m_s-m_u)$, $\gamma=2.148$ and $\Lambda_0=300\ MeV$ and $C$ and $D$ are
two new form factors.


We now evaluate the numerical values for the TPCs in this simple power expansion scenario in Eq. (\ref{SS}).
Our results   of $A_T$
($\bar A_T$) and ${\cal A}_T=(A_T-\bar{A}_T)/2$ are shown in
Table \ref{ATtable1}.
\begin{table}[h]
\tbl{Triple product correlation asymmetries (in percent)
of $A_T$ ($\bar A_T$) for $\bar{B}^0\to\Lambda\bar{p}\pi^+$
(${B}^0\to\bar{\Lambda} p\pi^-$) and ${\cal
A}_T=(A_T-\bar{A}_T)/2$.
}
{\begin{tabular}{|c|c|c|}
\hline
$A_T,\bar{A}_T,{\cal A}_T$&$\gamma =57^\circ$&$\gamma =0^\circ$\\
\hline\hline
$\delta\neq 0$&$12.0,\ -8.4,\ 10.2$&$2.0,\ 2.0,\ 0$\\
\hline
$\delta=0$&$10.4, -10.4, 10.4$&$0,\ \ 0,\ \ 0$\\
 \hline
\end{tabular}}
\label{ATtable1}
\end{table}
 As an illustration, in the table
we have also turned off the strong phase ($\delta=0$) by
taking the imaginary parts of the quark-loop rescattering effects
to be zero.  From the table, we see explicitly that ${\cal A}_T$ is indeed
nonzero and maximal in the absence of the strong phase. 
We note
that in our calculations we have neglected the final state
interactions due to electromagnetic and strong interactions, which are believed to be
small in three-body charmless baryonic decays \cite{GH05,ChengQCD}. 
We also note that  $A_{CP}\sim 1.1\%$ 
in $\bar{B}^0(B^0)\to\Lambda\bar{p}\pi^\pm$ 
can be induced but it is too small to be measured. 
It is
interesting to point out that in order to observe $A_T$ ($\bar{A}_T$) in
$\bar B^0 \to \Lambda\bar p\pi^+$ ($B^0\to \bar\Lambda p \pi^-$)
being at $12-8$\%, we need to have about $(1-2)\times 10^{8}$
$B\bar B$ pairs at $2\sigma$ level. This is within the reach of
the present day $B$ factories at KEK and SLAC and others that
would come up. It is clear that an experimental measurement of ${\cal A}_T$ 
is a reliable test of the CKM mechanism of CP violation and, moreover, 
it could be the first evidence of the direct T violation in B decays.
Finally, we remark that we  have also explored the direct CP and T violation in
 $B\to\Lambda\bar{\Lambda}K$ \cite{GH05} and we have found that
 the direct CP violating effect is  small but the T violating one is as large as
 that in  $B\to \Lambda \bar p \pi$.

In summary, 
we have shown that
the direct CP violating asymmetry in
$B^\pm\to p\bar p K^{*\pm}$ is around 22$\%$
and  the direct $T$ violating asymmetry in
$\bar B^0 \ra \Lambda \bar p \pi^+$ 
can be as large as
$12\%$, which are accessible to the current B factories at KEK
and SLAC as well as the future ones such as SuperB and LHCb.

\section*{Acknowledgements} This work is financially supported by
the National Science Council of Republic of China under the
contract \#: NSC-95-2112-M-007-059-MY3.


\begin{thebibliography}{99}
\bibitem{pdg} W.~M.~Yao {\it et al.}  [Particle Data Group], J.\ Phys.\ G {\bf 33}, 1 (2006).

\bibitem{CKM} N. Cabibbo, Phys. Rev. Lett. {\bf 10}, 531 (1963);
M. Kobayashi and K. Maskawa, Prog. Theor. Phys. {\bf 49}, 652
(1973).

\bibitem{CPT}C.~Q.~Geng and L.~Geng,
 Mod.\ Phys.\ Lett.\  A {\bf 22}, 651 (2007).
  
\bibitem{KT}G. Belanger and C. Q. Geng, \pr  {\bf D44}, 2789
(1991); P. Agrawal ${\it et\;al}$., \prl {\bf 67}, 537 (1991);
\pr {\bf D45}, 2383 (1992).

  
\bibitem{Hamiltonian}
Y.~H.~Chen, H.~Y.~Cheng, B.~Tseng and K.~C.~Yang, Phys.\ Rev. {\bf
D60}, 094014 (1999);
H.~Y.~Cheng and K.~C.~Yang, Phys.\ Rev. {\bf D62}, 054029 (2000).

\bibitem{ChuaHouTsai2}
C.~K.~Chua, W.~S.~Hou and S.~Y.~Tsai, Phys.\ Rev. {\bf D66}, 054004 (2002);
H.~Y.~Cheng and K.~C.~Yang, Phys.\ Rev. {\bf D66}, 014020 (2002);
S. Y. Tsai, ``Study of Three-body Baryonic B
Decays'', Ph. D thesis, National Taiwan University (2005).

\bibitem{GH05} C. Q. Geng and Y. K. Hsiao, Phys.\ Lett.\ B {\bf 619}, 305 (2005).

\bibitem{AngdisppK}
C.Q. Geng and Y.K. Hsiao, Phys.\ Rev.\  D {\bf 74}, 094023 (2006).
\bibitem{ppKstar}
C.~Q.~Geng, Y.~K.~Hsiao and J.~N.~Ng,
Phys.\ Rev.\ Lett.\  {\bf 98}, 011801 (2007);
 Phys.\ Rev.\  D {\bf 75}, 094013 (2007).
\bibitem{NF} C.Q. Geng and Y.K. Hsiao, Phys.\ Rev.\  D {\bf 75}, 094005 (2007).

\bibitem{CPppK}
B.~Aubert {\it et al.} [BABAR Collaboration], Phys.\ Rev. {\bf D72}, 051101 (2005);
Phys.\ Rev.\  D {\bf 76}, 092004 (2007);
see also T.B. Hryn'ova, ``Study of B Meson Decays to $p\bar{p}h$ Final
States'', Ph.D. thesis, Stanford University (2006).


\bibitem{Wang1} M.Z. Wang {\it et al.} [BELLE Collaboration], Phys. Rev. Lett. {\bf92}, 131801 (2004);
arXiv:0706.4167 [hep-ex].

\bibitem{GHT1} C. Q. Geng and Y. K. Hsiao, Phys.\ Rev.\ D {\bf 72}, 037901 (2005);
  Int.\ J.\ Mod.\ Phys.\  A {\bf 21}, 897 (2006).

\bibitem{belle0}
M.~Z.~Wang {\it et al.}  [Belle Collaboration],
  Phys.\ Rev.\  D {\bf 76}, 052004 (2007);
B.~Aubert  [BABAR Collaboration],
  arXiv:hep-ex/0608020.


\bibitem{soni}W.S. Hou and A. Soni, \prl  ${\bf 86}$, 4247 (2001).
\bibitem{chua} C.K. Chua, W.S. Hou and S.Y. Tsai, \pl ${\bf B 528}$, 233 (2002);
C.~K.~Chua and W.~S.~Hou, Eur.\ Phys.\ J.\ C{\bf 29}, 27
(2003).
\bibitem{group} F. Piccinini and A. D. Polosa, \pr
${\bf D 65}$,  097508 (2002); H. Y. Cheng and K.C. Yang, \pl ${\bf B
533}$, 271 (2002); 
 C. Q. Geng and Y. K. Hsiao, Phys.\ Lett.\ B {\bf 610}, 67 (2005).

 
 \bibitem{MS} D. Melikhov and B. Stech, Phys. Rev. {\bf D62} (2000)
014006.

\bibitem{Brodsky1}
G.~P.~Lepage and S.~J.~Brodsky, Phys.\ Rev.\ Lett.\  {\bf 43},
545(1979) [Erratum-ibid.\  {\bf 43}, 1625 (1979)];
G.~P.~Lepage and S.~J.~Brodsky, Phys.\ Rev.\ {\bf D22}, 2157 (1980);
S.~J.~Brodsky, G.~P.~Lepage and S.~A.~A.~Zaidi, Phys.\ Rev.\ {\bf D23}, 1152 (1981).

\bibitem{ChengQCD}H.Y. Cheng, 
 Int.\ J.\ Mod.\ Phys.\  A {\bf 21}, 650 (2006).
 
\bibitem{GH-f}
C. Q. Geng and Y. K. Hsiao, in preparation.



\end{thebibliography}
\end{document}